# MRI Investigations of Particle Motion within a Three-Dimensional Vibro-Fluidized Granular Bed

Mick D. Mantle[1*], Andrew J. Sederman[1], Lynn F. Gladden[1],

Jonathan M. Huntley[2], Tom W. Martin[2] Ricky D. Wildman[2]

Mark D. Shattuck[3]

(1) University of Cambridge, Department of Chemical Engineering, Pembroke Street, Cambridge, CB2 3RA, United Kingdom.

(2) Loughborough University, Wolfson School of Mechanical and Manufacturing Engineering, Loughborough LE11 3TU, Loughborough, United Kingdom.

(3) City College of New York, Levich Institute, 140th and Convent Ave., New York, NY 10031.

[*]To whom Correspondence should be addressed:

Dr M.D. Mantle
University of Cambridge
Department of Chemical Engineering
Pembroke Street,
Cambridge CB2 3RA,
United Kingdom.

Email: mdm20@cam.ac.uk

**Abstract**

The unique ability of magnetic resonance imaging (MRI) to provide spatial and temporal information from optically opaque systems, in three dimensions, make it an ideal tool to study the internal motion of rapid granular flows. This paper will focus on the use of ultra-fast velocity compensated MRI measurements to study particle velocity and density distributions in a granular gas, produced by vibrating vertically a bed of mustard seeds at 40 Hz. Specifically, a velocity compensated, double spin-echo, triggered, one-dimensional MRI profiling pulse sequence was developed. This gives an MRI temporal resolution of approximately 2 ms and also minimises MRI velocity artefacts. 12 phase measurements per vibration cycle were used. The data can be used to extract values of the mustard seed average velocity and velocity *propagators* (probability distributions functions) as a function of the phase of the vibration cycle and vertical height within the cell. The data show strong transient effects during the impact phase of the vibration. A detailed discussion of the temporal passage of the individual phase resolved, height resolved velocity distributions, along with seed velocity propagators at a fix height from the vibrating base is presented.

**Keywords:    Magnetic resonance imaging; granular flows; velocity distributions; vibro-fluidised beds.**

# 1 Introduction.

The use of solid materials in granular form is widespread in a variety of industries. Often the solid material is in the form of granules which have to be stored, transported and processed. However the behaviour of the granular material during these intermediate stages between material production and final use is often neglected. Whilst it is generally true that there is a great deal of information regarding specific granular systems, a more general approach that allows us to describe the behaviour of granular materials would be extremely beneficial. The dynamics of rapidly flowing granular materials has been developed from the theories associated with the more common states of matter such as gases and liquids. In particular, the microscopic kinetic theory of gases has been used as a basis for granular dynamics with the important distinction that the collisions in our macroscopic granular gas are energy dissipative and thus require an energy input to the granular system to sustain a dynamic or fluidised behaviour.

Whilst there is a large amount of theoretical literature describing granular dynamics most of the experimental studies concern *two-dimensional* granular motion which has been predominantly studied by high speed photography and subsequent image analysis [1-3]. Three-dimensional granular motion has been probed using positron emission particle tracking [4,5] and the results have shown reasonable agreement with theoretical predictions in relatively dilute systems. In recent years, nuclear magnetic resonance imaging (NMRI) has proved to be a useful non-invasive tool to probe granular dynamics. The interested reader is referred to a review by Mantle and Sederman [6] detailing the applications of MRI to granular systems. Much of the early MRI work in the 1990s focused on the use of poppy and/or mustard seeds because the cores of such materials have liquid like properties which allows them to be studied with MRI techniques. Whilst there are a number of experimental MRI studies of granular materials in horizontal systems, to our knowledge, there are only three MRI related reports in the literature concerning the continuous three-dimensional motion of particles in vertically vibrated systems [7-9]. Caprihan et al. used two-dimensional gradient echo imaging to study the motion of oil filled 2.0 mm spherical plastic beads in a vibrating bed. Period doubling and a band of high shear was detected using this technique. Most recently, Huan et al.[9] in a more comprehensive study of their earlier work [8] used the pulsed field gradient stimulated echo (PFG-STE) technique, at low external magnetic fields strengths ($B_0$ = 0.94 T), to probe the motion of particles in a vertically vibrated bed of poppy seed. A number of conclusions were drawn and a good agreement between the experimental PFG-STE data and the predictions of hydrodynamic theory were seen. In particular MRI was able to detect *the temperature inversion* phenomenon predicted by some hydrodynamic models.

The system under investigation here is similar to that of Huan et al. [9] in that the fluidised state is created by the vertical vibration of a container holding the granular material. We too have used mustard seeds as our granular material, but we are operating at a moderately high polarising magnetic field ($B_0$ = 7.07 T) which allows us to produce, for the first time, phase resolved velocity propagators (probability distribution functions) using a novel, one-dimensional pulsed field gradient double spin-echo technique that gives velocity compensation of the spatial (readout) gradients and negates the effect of mustard seed motion in linear background magnetic field gradients. The purpose of this paper is primarily to explain the nature of the measurement and describe qualitatively some of the results obtained. Further quantitative analysis of the data obtained from this system can be found elsewhere[10].

## 2  Experimental

*Vibro-fluidized bed*
For the present work we used mustard seeds on account of the relatively large diameter and narrow size range ($d$ = 2.04 ± 0.23 mm), and relatively small deviations from spherical geometry (aspect ratio = 1.17 ± 0.09). The mean grain mass was 5.66 mg. The cell containing the grains was machined from a permanently anti-static acetal co-polymer to reduce electrostatic charging effects, with a glass disc insert cell at the bottom. The internal radius of $R$ = 9 mm was constrained by the size of the NMR spectrometer bore, and the height was sufficient to prevent collisions with the top wall. The cell was mounted on top of a glass rod within the vertical bore; the rod was vibrated vertically by means of a camshaft and separate aluminium horizontal drive shaft attached to a dc electric motor., though the resulting resonances reduced the maximum practical drive frequency, $f$, to ~ 40 Hz, somewhat lower than the value of 50 Hz typically used in experimental studies using electromagnetic shakers[5]. The amplitude $A_0$ of the vibration cycle was in the range 1.25 mm and the number of mustard seeds used, $N_g$, was 55 corresponding to a monolayer of seeds. The cell was evacuated to reduce air drag on the grains.

*NMR acquisition*

A double spin-echo velocity profiling technique was developed (see figure 1) that allows velocity distributions within the three-dimensional cell to be measured as a function of both vertical position ($z$) and vibration phase (characterized by time $t$ after the trigger pulse). We chose here the vertical velocity component ($v_z$) since the granular temperature is normally highest in this direction, although other components could be measured equally easily at the expense of additional acquisition time. All MRI experiments were acquired on a Bruker Biospin DMX 300 spectrometer operating at a $^1$H frequency of 300.13 MHz. Spatial resolution was achieved using a three-orthogonal axis gradient system capable of producing a maximum gradient strength of 1 T m$^{-1}$. The field of view in the Z-direction was 50.0 mm and the number of data points acquired was 128, thereby giving an axial pixel resolution of 390 μm. A 25 mm $^1$H birdcage resonator was used to excite and detect the magnetisation from the mustard seeds and the $^1$H 90 degree pulse length was 32 μs. It should also be pointed out that ramped gradient were used throughout to minimise magnetic field Eddy currents (and hence extra spin dephasing) due to gradient switching. The gradient ramp-up and ramp-down times were 100 μs. The total echo time TE was 2.46 ms. Velocity encoding was achieved by using gradient 64 equal gradient increments from –0.6 to + 0.6 T m$^{-1}$. The length of the velocity encoding gradient, δ, was 580 μs. The delay between velocity encoding gradient pulses, Δ, was 1.40 ms. The NMR signal excitation was triggered at a fixed phase of the sample vibration. 12 increments in the vibration phase were used to sample one complete vibration cycle. The phase increments were calculated by dividing the inverse of the vibration frequency by 12. 8 scans, at a recycle time of 365 ms, were added to obtain a sufficient signal-to-noise ratio. Hence the total experimental time for a single vibration frequency was approximately 38 minutes. Following acquisition, the raw data was zero filled to 256 data points in the velocity encode dimension, and then a 2-dimensional Fourier transform was applied to each of the 12 data sets to give spatially encoded, vibration phase dependent, velocity profiles. These parameters enable us to determine velocities within a range $-0.95$ m s$^{-1}$ to $+0.95$ m s$^{-1}$ and with a resolution of 0.007 m s$^{-1}$. A unique feature of the data presented here is that the $^1$H background signal resulting from the acetal co-polymer sample holder, provides a useful reference for both the phase of the vibration cycle and an internal reference for the velocity.

## 3  Results and discussion.

*The MRI pulse scheme*

Dynamic MRI studies of transient systems can be inherently difficult to quantify because of MR signal loss/attenuation due to: (a) seed motion in so-called 'background' magnetic field gradients resulting from the different physical phases present- in this paper mustard seeds and air/vacuum; (b) seed motion during any non-MR-velocity encoding magnetic field gradients, i.e. those that are used to give spatial resolution, typically known as *readout* gradients. A further complication is that these two effects scale with the magnitude of the static polarising external magnetic field ($B_0$) used, which in our case is 7.07 Tesla. This is one reason why previous investigations on similar systems have been carried out on much smaller polarising static magnetic fields. However, the main advantage of using a stronger static polarising field is that more 'signal' is obtained per unit mass of substance, thereby yielding better signal-to-noise ratio and allowing, as in this case, spatially resolved, phase resolved, velocity distributions within a reasonable timescale. For these reasons we have developed an MR protocol that compensates for mustard seed signal loss due to motion in background magnetic field gradients and signal loss due to motion during the spatial readout gradient. Figure 1 shows the MR pulse sequence developed for this study. The exact details for the pulse sequence shown in figure 1 will be published elsewhere[11] but a brief description of its salient features is given below. The pulse sequence consists of three radio frequency (RF) pulses: the first $90^o$ RF pulse produces transverse magnetisation from the mustard seeds; the next two $180^o$ RF pulses form a standard double-spin-echo which compensates for any signal loss due to spin-spin ($T_2$) relaxation. However, the double spin-echo also compensates for any signal loss during linear seed motion in a linear magnetic field background gradient. The second $180^o$ pulse effectively reverses and hence negates any signal loss that occurred as a result of seed motion in the background gradient between the first ($90^o$) and second ($180^o$) RF pulse. In order to achieve spatial resolution of the mustard seeds in the vertical (z) direction with velocity compensation, the readout gradient ($G_{read}$) takes the form shown in figure 1. It can be shown[6] that any seed signal loss, due to linear motion of the seeds during the $G_{read}$ gradients, is made equal to zero by ensuring that the first moment of the $G_{read}$ gradients is zero.

*Data interpretation*

Figure 2 shows the mustard seed probability distributions, $S(v_z, z, t)$, corresponding to the third of the 12 phases acquired within a single vibration cycle of mustard seeds ($A_0$=1.25 mm, $f$ = 31.8 Hz, $N_g$= 55) with velocity along the vertical axis, and position in the z-direction along the horizontal axis. The container wall and base of the container can clearly be seen on the plot, the latter providing a convenient marker for the origin of the z-axis, defined as the mean location of the top, i.e. impacting surface of the glass plate. At this particular phase in the vibration cycle the distribution of velocities has a strong z-dependence and is clearly irregular over the vertical distance. Figure 2 shows that some of the particles are moving towards the base, as indicated by negative $v_z$ velocity values, and some particles are moving away from the base (positive velocity) against the influence of gravity. Figure 3 shows the complete passage of the same granular system during one vibration cycle. In this figure the contribution from the acetal co-polymer cell has been removed from the images for clarity, and the base of the cell has been binary gated so that it appears as pure white pixels in the series of images. Following the time series of velocity against height distributions from frame 2 to frame 5 we see that the base of the cell is moving down and that the velocity of the seeds transforms from a predominantly positive upwards motion in frame 2 to a predominantly downwards motion in frames 5, 6 and 7. Between frames 5 and 6 we see that the magnitude of the negative z-velocity of the base decreases as the base decelerates. At frame 7 the base is on the cusp of going from a downward motion to an upward motion and hence the velocity appears to be nearly zero; but note, nearly all the seeds are now travelling with a downward negative velocity. Frame 8 indicates that the base is now moving upwards towards the free falling seeds, as indicated by the majority of the seeds having a negative velocity. Impact of the bottom of the base with the seeds occurs between frames 8 and 9 as evidenced by the sudden appearance of high positive velocities (circled in frame 9 of figure 3) with an average velocity of approximately +0.5 m s$^{-1}$. From frames 10, through 12, and back to 1, we see that the base of the cell is still moving with a positive velocity. During theses frames, more of the seeds come into contact with the base and thus receive a positive impulse. This interpretation is supported by the fact the circled region in frames 10, 11, and 12 grows in size along the horizontal axis and has a greater greyscale intensity indicative of a greater seed population; the average velocity decreases to a value of approximately +0.2 m s$^{-1}$ in frame 12. Interestingly, in frame 1 there is evidence for a second region of seeds with a distinct high velocity which is suggestive of a second 'shock' or impulse to the system. Note that the base of the cell in frame 1 is still moving with a positive upwards velocity.

*Seed velocity propagators*

Figure 4 shows phase resolved, seed velocity probability distributions (propagators) taken at a fixed height of 7.5 mm above the average position of the vibrating base. The best fit Gaussians, equivalent to Maxwellian velocity distributions, are also plotted for comparison. Taking each propagator plot on each individual phase-plot basis, we see that for plots 1 and 2, the average velocity of the seeds at this fixed height is approximately +0.15 and +0.10 m s$^{-1}$ respectively. Plot 3 shows that the average velocity of the seeds is approximately zero, implying nearly equal amounts of seeds moving towards and away from the base. Plots 4 to 10 all have negative average velocities of in the range –0.04 m s$^{-1}$ to -0.15 m s$^{-1}$. Given the rapid cyclical changes in mean velocity, plots 1 thorough 10 are all fit reasonably well by a Gaussian form. However, plots 11 and 12 indicate a significant deviation from a single Gaussian and are likely to be better fitted using a bimodal distribution.  Figure 5 shows the data from figure 4, averaged over the 12 individual phases, to give a single, non-phase resolved velocity propagator, along with a corresponding best fit Gaussian.  Figure 5 shows that the non-phase resolved propagator deviates significantly from the Gaussian form and underlines the importance of a phase resolved propagator measurement, in order to fully appreciate some of the mechanisms present during granular fluidization by a vertical vibrating base.  It would be virtually impossible to extract the kinematic subtleties occurring within a single cycle using conventional phase-unresolved MRI measurements.

## 4 Conclusions

A three-dimensional vibrofluidised bed has been studied *in-situ* using a specially designed MRI pulse sequence to reduce signal losses due to coherent motion in both spatial encoding gradients and background magnetic field gradients in a high field MRI spectrometer.  The velocity probability distribution functions in the vertical direction (obtained from a two-dimensional Fourier transform of the raw data) during a complete vibration cycle show both Maxwellian behavior and a bi-modal Maxwellian behavior, the latter being more pronounced immediately after the compressional phase of the vibration. A free falling behavior, characterized by a much more Gaussian like propagator for the whole of the medium, is subsequently seen in the following dilation phase.  The phase resolved velocity propagators extracted at a specific height from within the vibrofluidised bed stress the importance of performing the measurement with cyclical phase resolution, in order to begin to be able to understand the underlying dynamics to this system.

## 5 Acknowledgements

MDM, AJS and LFG wish to thank the EPSRC for the provision of the MRI spectrometer.  JMH is also grateful to the Royal Society and Wolfson Foundation for a Royal Society-Wolfson Foundation Merit Award.

**Figure Captions**

Figure 1.  The double spin-echo velocity compensated magnetic resonance pulse sequence used in this study.

Figure 2.  Plot showing the probability distribution $S(v_z, z, t)$ of $v_z$-velocity variation of the mustard seeds (horizontal axis) against the displacement in the z-direction (vertical axis) for one stage in the vibration cycle.  The colour bar represents signal strength (seed population) in arbitrary units.

Figure 3.  As figure two except that all 12 sampling points of a single vibrational cycle are displayed as individual frames numbered 1 to 12.  The time interval between each successive frame is approximately 2 ms.

Figure 4.  The individual $v_z$ velocity propagators (solid line) from a data set at a fixed height of $z$ = 7.5 mm for each of 12-phase frames with $A_0$=1.25mm, f = 31.8 Hz and $N_g$ = 55.  Dotted lines represent best fit Gaussians.

Figure 5.  Non-phase resolved velocity propagator (solid line) for the data in figure 4.  The velocity distribution is obtained by averaged the 12 individual phases shown in figure 4.  The dotted line shows the best fit Gaussian.

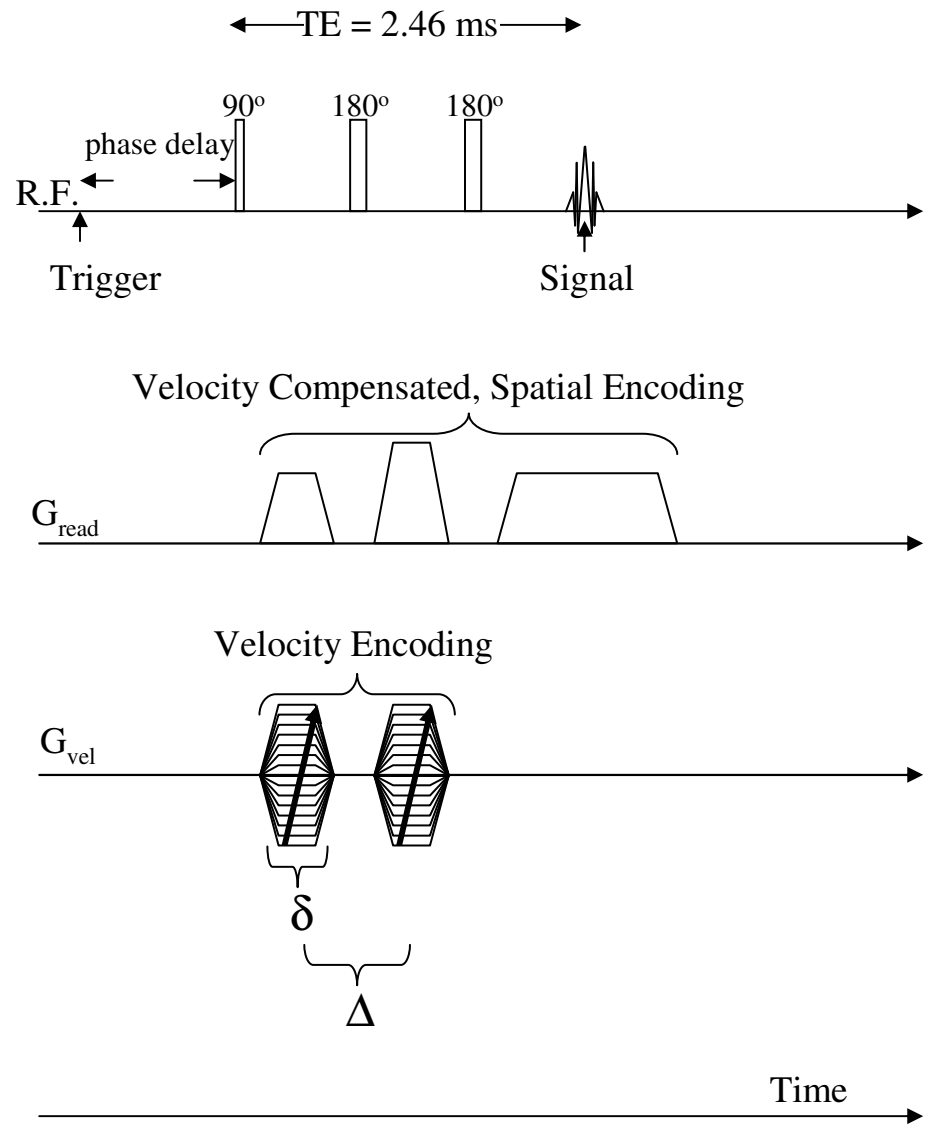

Figure 1

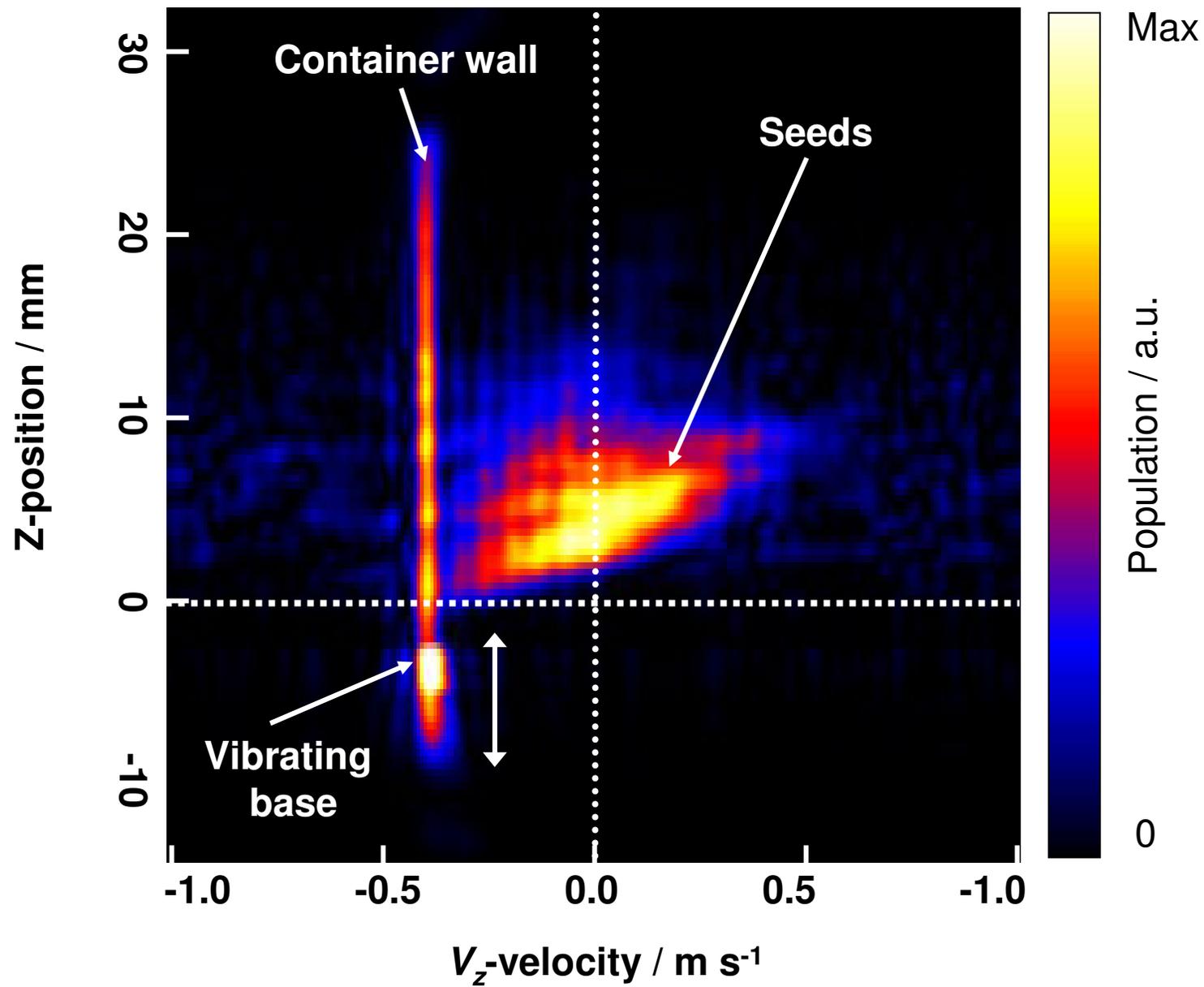

Figure 2

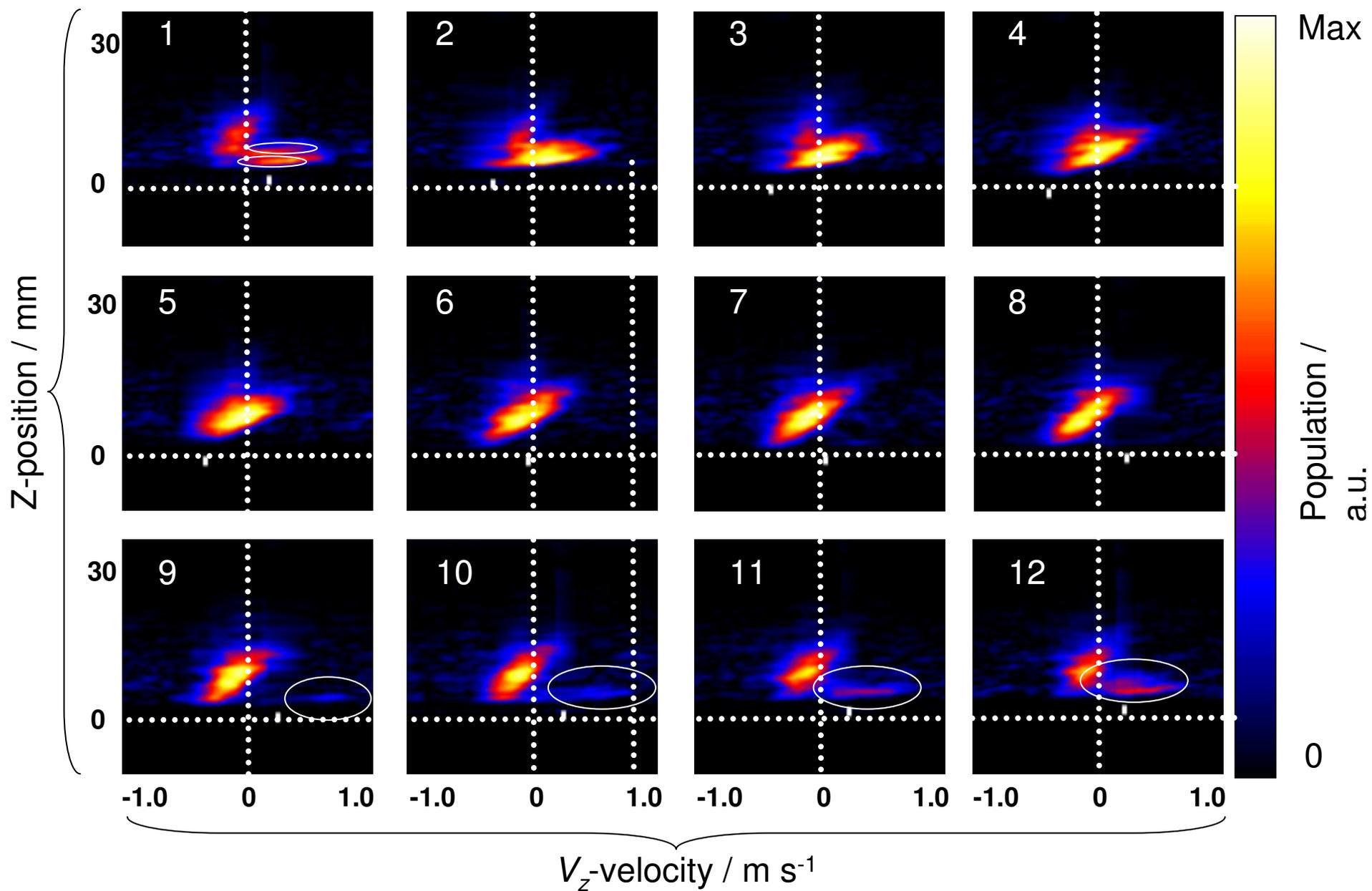

Figure 3

Figure 4

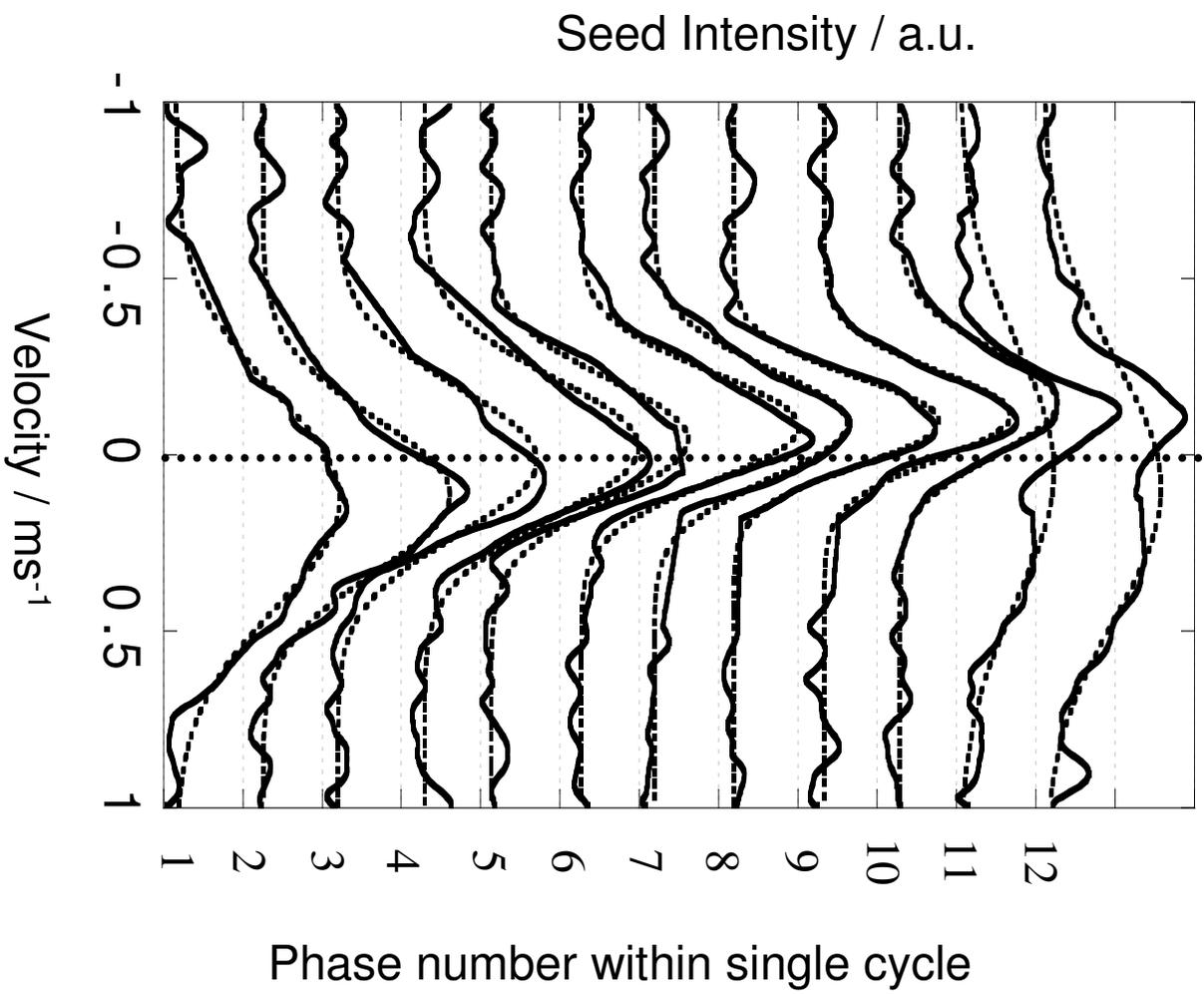

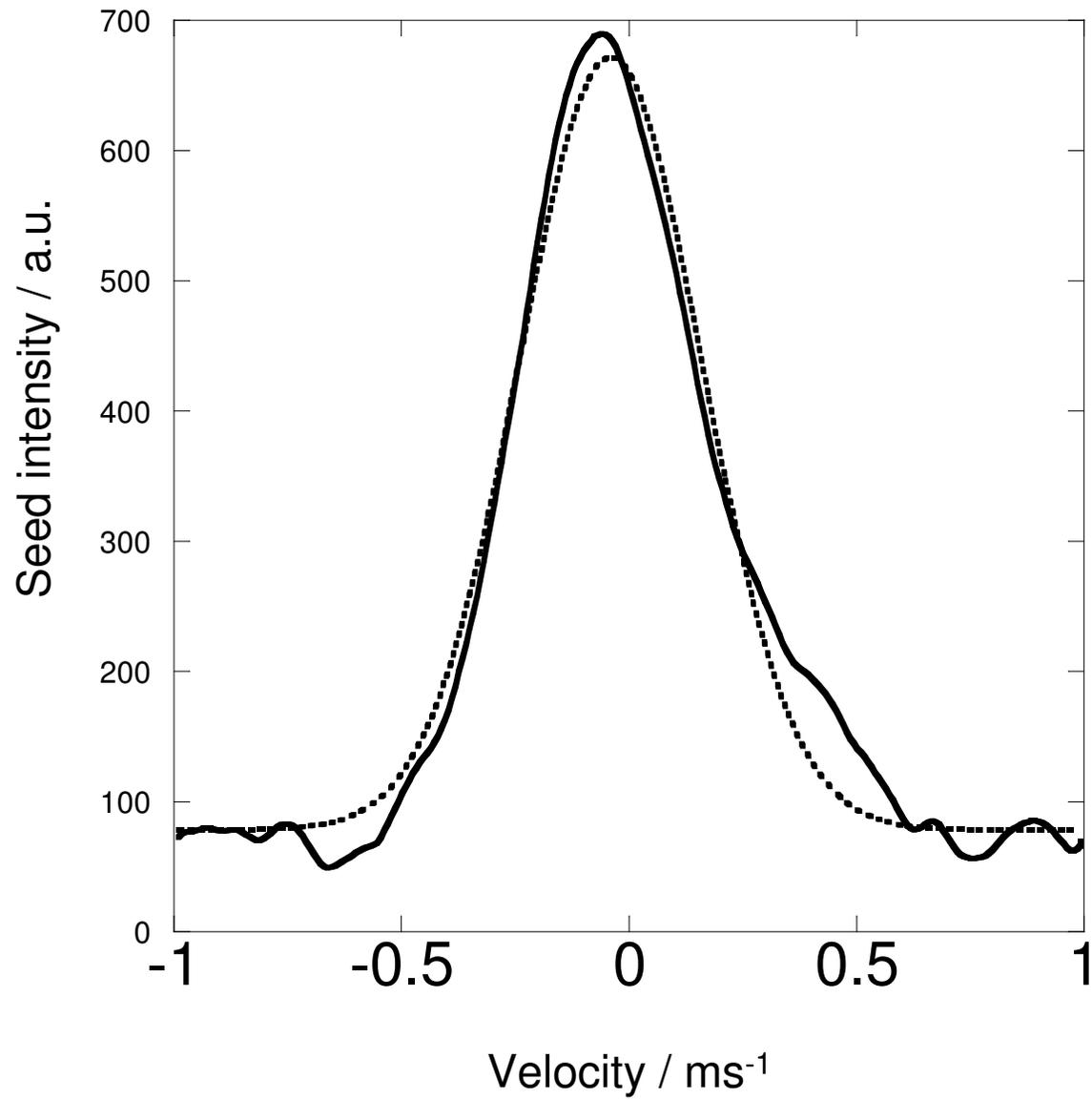

Figure 5